\documentclass[10pt]{wlscirep}
\usepackage{graphicx}
\usepackage{caption}
\usepackage{subcaption}
\usepackage{authblk}
\title{The Multi-mode Acoustic Gravitational Wave Experiment: MAGE}
\author{William M. Campbell, Maxim Goryachev and Michael E. Tobar}
\affil{ARC Centre of Excellence for Engineered Quantum Systems and ARC Centre of Excellence for Dark Matter Particle Physics, Department of Physics, University of Western Australia, 35 Stirling Highway, Crawley WA 6009, Australia}
\begin{abstract}
The Multi-mode Acoustic Gravitational wave Experiment (MAGE) is a high frequency gravitational wave detection experiment. In its first stage, the experiment features two near-identical quartz bulk acoustic wave resonators that act as strain antennas with spectral sensitivity as low as $6.6\times 10^{-21} \left[\textrm{strain}\right]/\sqrt{\textrm{Hz}}$ in multiple narrow bands across MHz frequencies. MAGE is the successor to the initial path-finding experiments; GEN 1 and GEN 2. These precursor runs demonstrated the successful use of the technology, employing a single quartz gravitational wave detector that found significantly strong and rare transient features. As the next step to this initial experiment, MAGE will employ further systematic rejection strategies by adding an additional quartz detector such that localised strains incident on just a single detector can be identified. The primary goals of MAGE will be to target signatures arising from objects and/or particles beyond that of the standard model, as well as identifying the source of the rare events seen in the predecessor experiment. The experimental set-up, current status and future directions for MAGE are discussed. Calibration procedures of the detector and signal amplification chain are presented. The sensitivity of MAGE to gravitational waves is estimated from knowledge of the quartz resonators. Finally, MAGE is assembled and tested in order to determine the thermal state of its new components.
\end{abstract}
\begin{document}

\flushbottom
\maketitle

\thispagestyle{empty}
\section*{Introduction}
The dawn of gravitational wave (GW) astronomy has arisen after the detection of numerous black hole and neutron star merger events by the current generation of interferometric GW detectors \cite{Abbott2015,Abbott2016,Abbott2017, Abbott2020,Acernese2014, Abe2022, Aassi2015}. However, such ground based detectors currently are only targeting gravitational signals that lie in the frequency range of 10 Hz-1 kHz. This limited spectrum is in contrast to that of electromagnetic astronomy, where observations may be conducted over a vast frequency range (from radio frequencies to X-ray). Broadening the scope of GW astronomy to include the pursuit of \textit{High Frequency} Gravitational Wave (HFGW) sources of 10 kHz and above is thus immediately recognised as a natural next step for GW astronomy.\\
While all currently known astrophysical sources of GWs lie in the nHz to kHz bands, the theoretical motivation for potential sources of HFGWs has seen a steady increase in recent times\cite{Aggarwal2021}. Proposed models which give rise to HFGWs in these higher frequency bands include, but are not limited to; various inflation scenarios\cite{Bartolo_2016}, early universe phase transitions\cite{Hindmarsh2015, Hindmarsh2017, Guo_2021}, cosmic strings\cite{Damour2000,Damour2001}, primordial black hole merger events\cite{Raidal_2019} and exotic compact objects \cite{Giudice_2016}. Such higher frequency sources present an additional interest in that any detection of gravitational radiation in this range would suggest the existence of beyond the standard model physics.\\
Many of the greatest challenges in physics today require solutions that aren't known to exist within the standard model of particle physics. Therefore, searching for any signatures beyond this model could provide critical insight into many other sectors of physics, such as the nature of dark matter and dark energy. Searching for HFGWs is thus a well motivated and interesting challenge, prompting the need for significant research into the requisite technology and experimentation that will allow detector sensitivity in these higher frequency bands.\\
\begin{figure}[ht]
\centering
\includegraphics[width = 0.5\textwidth]{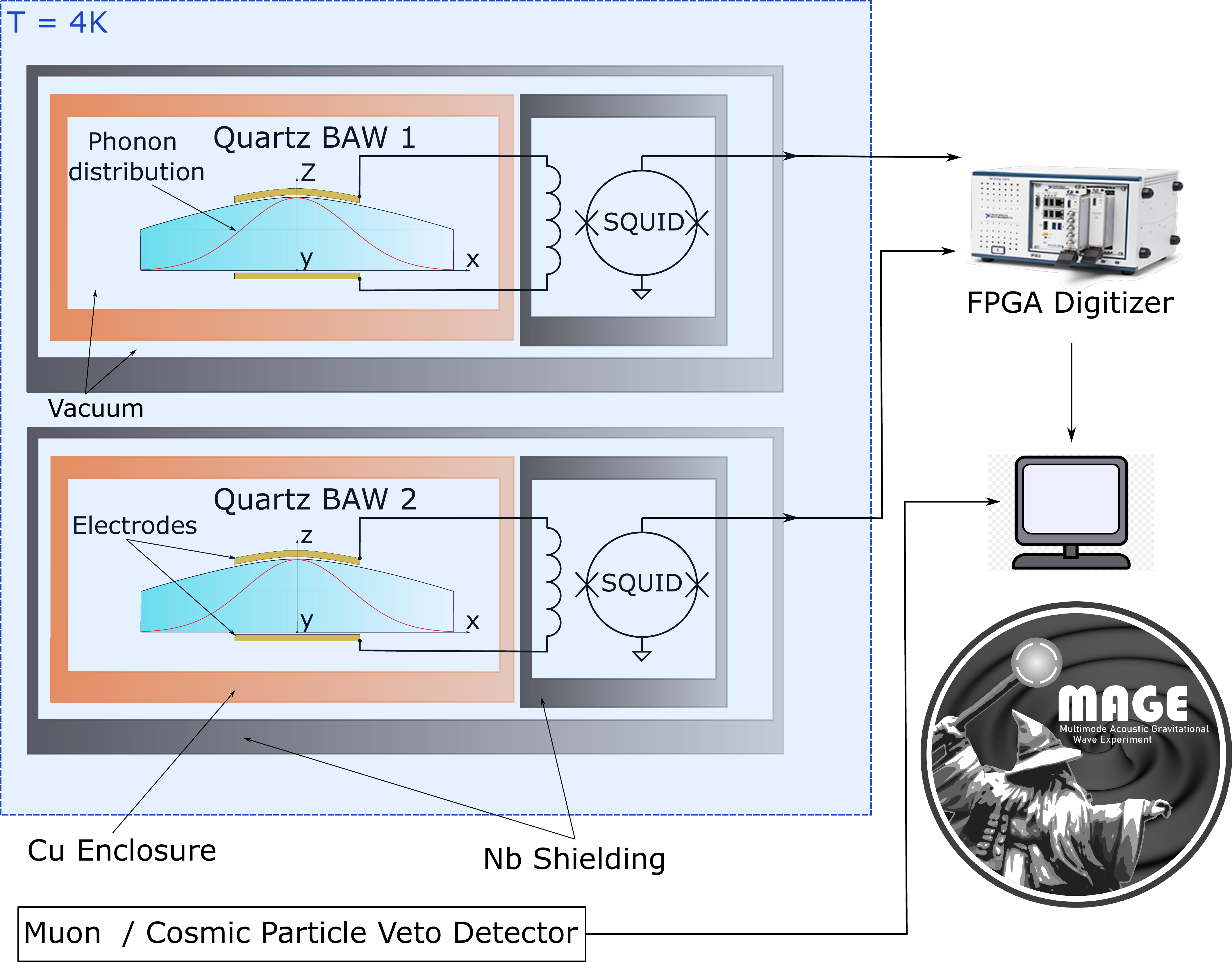}
\caption{\label{fig:MAGEschematic}Schematic of the proposed MAGE experiment. It features two (near) identical quartz BAW HFGW detectors read out across multiple channels using a high - throughput FPGA based digitizer. An external scintillator will also be utilised for cosmic muon veto.}
\end{figure}
For Michelson interferometric detectors such as the current second generation and planned third generation ensemble, the main challenge prohibiting sensitivity at higher frequencies is a violation of the long wave length limit for HFGWs incident upon the km scale long arms of the interferometer. While there are some interferometric detectors \cite{Martinez2020} that feature much shorter arm lengths and are thus sensitive to GWs greater than 10kHz, they still suffer from the fundamental limitation that shorter interferometer arm length reduces the resulting strain signal from impinging GWs in the detector. Due to this fundamental trade-off, other solutions may be more optimal than interferometry for the case of HFGW detection.\\
Thankfully other solutions have been developed \cite{Cruise2006, Npishizawa2008, Arvanitaki2015} including some recent proposals \cite{Ito2020, Ejlli2019} further signifying an increasing motivation for HFGW detection. Additionally, many new technologies that are being developed to primarily search for dark matter have also been shown to exhibit sensitivity to HFGWs \cite{Tobar2022, Domcke2022}. One of the most mature of these new HFGW solutions in terms of experimental development is a revival of the resonant-mass GW antenna which was the dominant style of experiment at the origin of GW detection \cite{Weber1966}. Recently \cite{Goryachev2014} it was proposed that utilising quartz crystal bulk acoustic wave resonators (BAWs), one can construct a resonant mass antenna that is highly sensitive to impinging strain fields at MHz frequencies. Such resonators are designed for precision frequency and metrology applications and have previously been utilised as highly sensitive probes of new physics \cite{Campbell2021,Goryachev2018}. In this proposal, it was shown that the incident strain fields of HFGWs couple to acoustic modes in the crystal bulk. Excitation of these bulk acoustic wave modes induces a charge density via the piezoelectric effect, which can then be read-out through small electrodes held close to the crystal surface. When subject to a cryogenic environment the thermally induced excitations of these modes are greatly reduced, such that the device acts as a highly sensitive external strain antenna. A previous work  \cite{Goryachev2014b} successfully showed that such thermal excitations at $T=4K$ could be directly observed by coupling the quartz resonator's electrodes to the input of a superconducting quantum interference device (SQUID) amplifier \cite{Clarke2006}. In an initial proof of concept experiment \cite{Goryachev2021} such a system in which a single quartz resonator coupled with a SQUID, was successfully operated as a sensitive HFGW detector. This experiment collected HFGW sensitive data in two phases referred to as GEN 1 and GEN 2.\\
This work introduces a further dedicated experiment, currently in development, to search for HFGWs based upon the quartz BAW architecture; the multi-mode acoustic gravitational wave experiment. MAGE is a well developed step forward in the growing community of HFGW detection, offering a relatively low cost solution on the order of \$180 thousand USD (including an estimated \$150 thousand USD for a 4 K cryogenic system).\\ 
In this work it is shown that MAGE is a currently active and well motivated HFGW detection experiment with competitive sensitivity that agrees well within estimation via the Nyquist noise model for the detector. The structure of this article is as follows; the first section consists of a summary of the predecessor GEN 1 and GEN 2 experiments, in which significant transient features where detected. This discussion provides context and motivates the successor experiment in MAGE. The following section details the experimental set-up of MAGE as well as its broader scope and capabilities. Calibration of MAGEs signal amplification chain is then presented in the following section as this procedure is critical in determining the experiments sensitivity to HFGWs. A subsequent estimation of MAGEs sensitivity to HFGWs based on a thermal noise model is then presented in order to highlight the significance of MAGE in the landscape of other HFGW detection experiments and potential HFGW sources. Finally, experimental results are presented in the last section in which the assembled MAGE detector's thermal state and sensitivity are directly determined based on the calibration measurements of the preceding section. The experimentally determined sensitivity can then be referenced against the estimations made via noise model. The results presented establish MAGE as an active, well understood and ongoing experiment with current sensitivity to HFGW sources.
\section*{The GEN 1 and GEN 2 Experiments}
The predecessor and proof of concept experiment to MAGE\cite{Goryachev2021} featured 153 days of intermittent data collection from April of 2019 to March of 2020, in which data taking was broken into two phases named GEN 1, and GEN 2. In both these experiments a single quartz BAW detector coupled to a SQUID amplifier was sensitive to HFGWs in two narrow frequency bands that correspond to two higher order overtone modes of the quartz BAW's fundamental acoustic resonances. The GEN 1 phase monitored two acoustic modes at 8.392 and 5.505 MHz, while the GEN 2 phase monitored modes at 4.993 and 5.505 MHz. When averaged across the entire operational run time, all modes displayed peak spectral strain sensitivities better than $1.6\times 10^{-20} \left[\textrm{strain}\right]/\sqrt{\textrm{Hz}}$. Comparing this result to the only other operational HFGW detector in the same frequency range; the Holometer experiment, we see this detector displays a broadband (1-13 MHz) sensitivity of at best  $2.4\times 10^{-19} \left[\textrm{strain}\right]/\sqrt{\textrm{Hz}}$. However, this multi-interferometric experiment can attain increased sensitivity beyond that of the GEN 1 and GEN 2 experiments via cross correlation of two interferometers.\\
Data acquisition for GEN 1 and 2 was achieved by filtering the voltage signal at the output of the SQUID via lock-in amplifiers referenced to the known frequencies of the quartz modes. The signal output by the lock-ins was then digitized and sampled at 100 Hz. A transient impulse search analysis was performed on the resulting data streams, in which any  larger than $5\sigma$ excitations of the detector's vibrational energy, which can be inferred from the output voltage signal, were searched for over a millisecond time scale. Two significantly strong and rare energy impulse features were detected. The first transient feature exhibited a strong increase to the oscillation amplitude of the 5.505 MHz mode during the GEN 1 phase, which decayed at a rate corresponding to the expected energy decay constant of the crystal mode. The second feature saw a similar impulse observed at 5.505 MHz in the GEN 2 phase, however a similar, slightly weaker impulse was also seen in the 4.993 MHz mode at the same instant. In this preceding work, possible sources for these features were discussed and some backgrounds were able to be excluded. It is largely believed by the GW community that the observed signals are far too strong in amplitude to be considered as HFGW events \cite{Domenech2021, Lasky2021}. However, in order to further characterise the systematics in the detector these observations must be further understood. Thus, to exclude non-gravitational energy impulse's such as crystal stress relaxations, radiative processes, and charged cosmic particles incident upon the detector, a next generation experiment with improved background veto systems is needed.
\section*{MAGE}
The MAGE experiment, as presented in figure \ref{fig:MAGEschematic}, further develops upon the GEN 1 and 2 experiments by  featuring two identical quartz BAW HFGW detectors, coupled to two independent SQUID amplifiers. These devices are placed inside a dilution refrigerator where they are held at a constant temperature of 4 K. Each detector consists of a piezoelectric BVA SC-cut \cite{besson2, stevens2013} plano-convex quartz plate 1mm thick and 30mm in diameter that sits in a copper enclosure held at a dedicated vacuum pressure. Two signal pins extend from each of these enclosures that electrically connect to small copper electrodes held closely to the vibrational surfaces the crystals. The pins of each resonator are coupled to the SQUID amplifiers via inductive input coils, and then each device is placed inside a dedicated Niobium cavity that becomes superconducting at T$~<10$ K. These superconducting shields in combination with various layers of Faraday shielding provided by the dilution fridge decouple both the detectors and SQUID sensors from stray electromagnetic fields. Local acoustic disturbances incident upon the fridge or detectors are also insignificant sources of noise for MAGE thanks to shielding by the internal vacuum environment as well as the high frequency range of operation.\\
MAGE features two identical quartz BAW detectors so that transient energy excitations of just a single detector can be excluded as potential HFGW events incident on both crystals via coincident analysis. This allows for the exclusion of observable background sources, such as internal crystal stress relaxation, radioactive decays and charged cosmic particles incident on a single detector. Each of these detectors is sensitive to gravitational radiation in multiple extremely narrow bands corresponding to the crystal's longitudinal overtone modes. For the case of the quartz crystals in question, these modes populate the MHz-GHz frequency band. Such modes exhibit uncommonly high quality factors of up to tens of billions at cryogenic temperatures \cite{Screp, galliou2011}, which leads to a fundamental constraint on the sensitive bandwidth of these detectors in that the bandwidth of these modes are usually $<$1 Hz. To combat this constraint, MAGE will simultaneously monitor multiple modes at once, making use of the crystals large mode density to effectively sample gravitational radiation in many narrow bands over an overarching broader spectrum. In practice\cite{Goryachev2021} this can be achieved by utilising lock-in amplification to mix down and amplify a small frequency window centred on a crystal mode.\\
In order to monitor a high number of modes at once across multiple detectors, MAGE will utilise a top of the range field programmable gate array (FPGA) based digitizer with reconfigurable customisation via integration with LabView software.  The rms amplitude of the voltage signal output by the SQUID amplifier typically lies in a $0.1-10~\mu$V range depending on the SQUID gain, and is easily coupled to the analogue input of the digitizer which specifies an input noise density as low as $\approx 9~n\mathrm{V}/\sqrt{\mathrm{Hz}}$. This hardware upgrade completely overhauls the data acquisition process for MAGE when compared to the predecessor experiments which featured two Stanford Research Systems SR844 lock-in amplifiers, giving MAGE a significant advantage over GEN 1 /2 in that it the data acquisition can be adapted and modified as hardware requirements change in the future. The digitizer has thus been programmed to take the 500 MHz sampled input signal and provide digital lock-in amplification in multiple parallel channels via digital signal synthesis and decimation. This will allow for multiple modes in each crystal to be continuously and simultaneously monitored thanks to the high throughput and memory of the on-board FPGA, which features 60 direct memory access channels, giving 30 lock-in channels for each crystal. The current iteration of MAGE thus has the capability to monitor 30 acoustic modes in each crystal at once.\\
This completes the introduction of MAGE in terms of its current features. However, further planned developments will increase the scope of MAGE as ongoing upgrades are implemented in parallel with HFGW-sensitive data acquisition. The most immediate of which is a muon veto system to be employed into the MAGE experiment at a later date. This detector would allow MAGE to exclude energy depositions in the detector sourced by charged cosmic particles from candidate HFGW signal events. Cosmic particles such as muons that collide with the detectors will cause energy depositions that can excite the acoustic modes of interest. These signals would be indistinguishable from the transient energy impulse caused by a passing gravitational wave. In order to veto such energy depositions, future iterations of MAGE will feature an externally placed scintillation detector with a large area, such solutions made from fabricated plastics have been demonstrated elsewhere \cite{yu2017,Adhikari2018}.\\
Further planned developments to MAGE include a software upgrade that packages several digital data streams into an individual FPGA memory channel, allowing for multiple modes to be monitored per channel. Such an upgrade would result in the maximum number of simultaneously observables crystal modes being limited purely by the total throughput bandwidth of the digitizer, of which the theoretical maximum is 16.375 Gb/s. To give a sense of scale the current digital lock in program uses $\approx 60 \%$ of the digitizers available memory resources, leaving sufficient room to increase the maximum attainable throughput through such software techniques.\\
Due to MAGE being situated inside a dilution refrigerator, it possess the ability to be pushed to lower thermal environments of $T=20$ mK. This would reduce the thermal noise of the detector and increase the overall sensitivity of MAGE. However, new SQUID amplifiers that can operate at these lower temperatures would need to be acquired, and the thermal behaviour of the SQUID-quartz coupled system would need to be investigated at mK temperatures. Thus, pushing MAGE into sub-Kelvin operation and greatly improving its sensitivity is an interesting prospect for future work.\\
In acquiring the hardware for MAGE, the quartz BAW detector, shielding cavity, and SQUID amplifier from the GEN 1 and GEN 2 experiments could be re-used in establishing one of the two detectors required. A second identical quartz BAW resonator, SQUID amplifier and dedicated Niobium shielding cavity where thus acquired to complete the construction of the second detector system. In order to complete the integration of the new hardware components to MAGE the second detector must be calibrated such that the gain properties of the signal read-out path is understood and agrees well with prediction. Thus, all reference in the following unless stated otherwise will be made to this second quartz resonator and SQUID amplifier, which differs from that used in the GEN 1 and GEN 2 experiments.\\
\begin{figure}
\centering
\includegraphics[width = 0.48\textwidth]{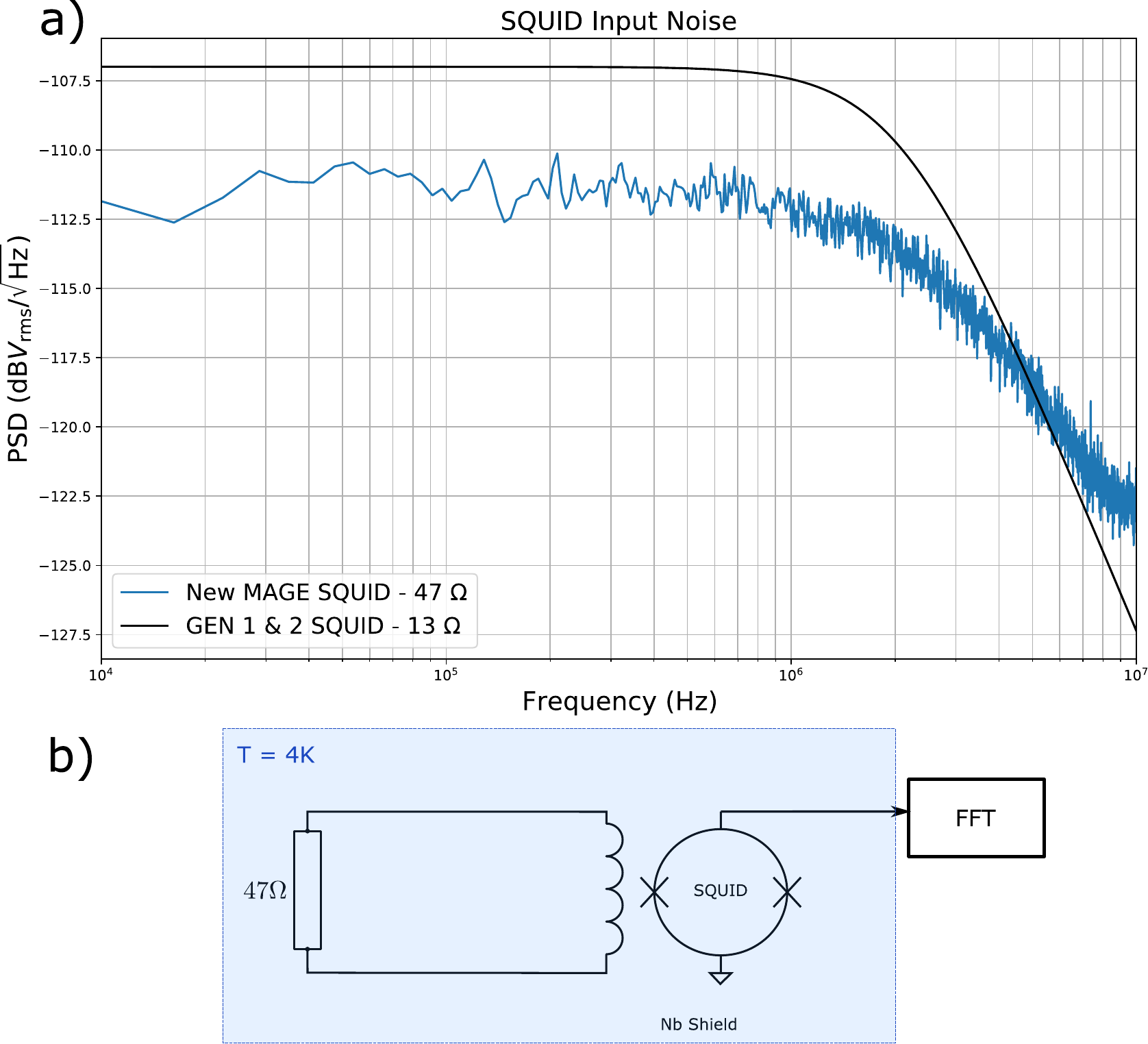}
\caption{a) Power spectral density of voltage fluctuations at the SQUID output coil (blue trace) and also the measured noise floor of the SQUID control electronics (orange trace), also plotted in black is the fit to the voltage spectrum obtained from another SQUID sensor utilised in the previous GEN 1 / GEN 2 HFGW detector. b) Diagram detailing the set-up for calibration, here `FFT' denotes a fast Fourier transform spectrum analyser.}
\label{fig:Vnoise}
\end{figure}
\section*{Calibration}
The SQUID sensor to be used in the second detector is a single stage current sensor from Magnicon. In order to calibrate its performance a temperature independent resistor of 47 $\Omega$ was shorted across the sensor's input coil. This configuration was placed inside a Niobium shield and mounted to the cold plate of a dilution refrigerator\cite{Zu2022} where it was cooled to 4 K. This refrigerator provides a large, stable and well isolated experimental space that is also shared by other well developed fundamental physics experiments such as the ORGAN dark matter detector\cite{Quiskamp2022}. A 3 metre long cryo-cable connects the SQUID sensor chip at the refrigerator cold plate to dedicated external control electronics at room temperature. The room temperature control stage for each SQUID amplifier provides feedback, signal readout and biasing operations to the SQUID sensors as well as a further room temperature gain stage. Many parameters that dictate these operations can be modified using control software, allowing for a large degree of tuning of the SQUID amplifier system. The spectral density of voltage fluctuations taken at the amplifiers output were then measured using a spectrum analyser, as shown in figure \ref{fig:Vnoise}. The dynamics of the output voltage noise is dominated by a high frequency roll off that is due to the bandwidth of the cryo-cable and room temperature amplifier. The net SQUID sensor and room temperature electronics can be considered an effective transimpedance amplifier characterised by $Z_\mathrm{SQUID}$ which refers the output voltage noise to the current fluctuations in the input coil. The following calculation can be made to determine $Z_\mathrm{SQUID}$ assuming a purely inductive input circuit model.
\begin{equation}\label{eqn:ZSQUID}
Z_\mathrm{SQUID}=\frac{\delta v_\mathrm{rms}^{(\mathrm{out})}}{\delta i_\mathrm{rms}^{(\mathrm{in})}} =\delta v_\mathrm{rms}^{(\mathrm{out})} \left(\frac{\sqrt{4k_bT_0R}}{\sqrt{R^2+(2\pi f L_i)^2}}\right)^{-1},
\end{equation}
where $R$ is the resistance of the input load, $k_b$ is Boltzman's constant $T_0$ is the system physical temperature, $L_i$ is the inductance of the input coil and $Z_\mathrm{SQUID}$ is the transimpedance gain. By finding an analytical fit to the spectrum of $\delta v_\mathrm{rms}^{(\mathrm{out})}$, equation (\ref{eqn:ZSQUID}) was used to determine $Z_\mathrm{SQUID} = 1.3\times10^6$ with a with a 3 dB bandwidth $f_\mathrm{3dB} = 2.4$ MHz. This represents an $18\%$ increase in low frequency gain and a $16\%$ increase bandwidth when compared to the SQUID used previously the GEN 1 and GEN 2 experiments. While this bandwidth is sufficient to amplify quartz modes at frequencies $f<15 MHz$, it ultimately limits the ability to utilise the higher frequency overtone modes of the crystal. Potential future hardware solutions to improve this bandwidth include using a shorter cryo-cable, or upgrading the room temperature amplifier and electronics altogether. However, for the current run of MAGE looking at modes within this bandwidth will be sufficient.
\section*{Sensitivity to HFGWs}
\begin{figure}
\centering
\includegraphics[width=0.45\textwidth]{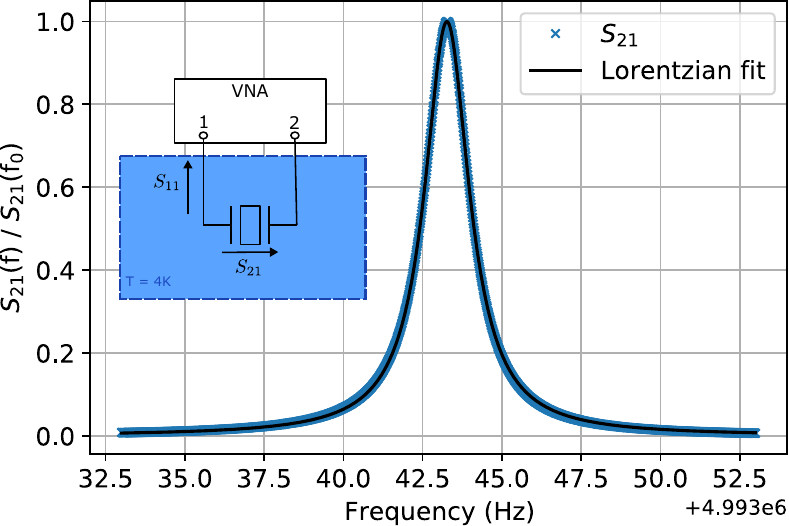}
\caption{\label{fig:Q}The measured S21 transmission coefficient of the quartz resonator's C$_{3,0,0}$ mode at 4.993 MHz is shown by the blue markers. The black trace shows a Lorentzian fit to the measured values. The values have been normalised by the maximum transmission on resonance $S_{21}$(f$_0$). Inset shows the set-up for the network analysis measurement scheme.}
\end{figure}
\begin{figure}
\centering
\includegraphics[width=0.48\textwidth]{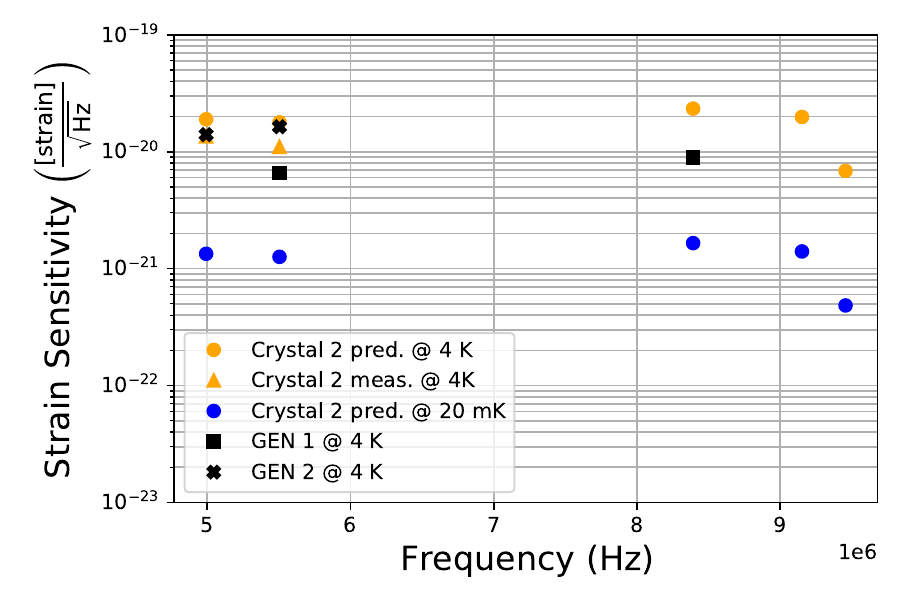}
\caption{\label{fig:strain}Peak spectral strain sensitivity for multiple quartz BAW modes. The predicted sensitivity from measuring $Q_\lambda$ of the resonator described in this work at 4 K, as recorded in table \ref{tab:results}, is plotted in the orange circles. The orange triangles show the sensitivity measured by the assembled detector. The black markers correspond to the measured sensitivity of the GEN 1 and GEN 2 experiments. The blue markers correspond to a predicted estimate for the sensitivity of a 20 mK device.} 
\end{figure}

The sensitivity of the second quartz detector to HFGWs in the vicinity of it's acoustic modes can be estimated with some knowledge of the device's thermal state. The single sided spectral strain sensitivity $S_h^+(\omega)$ of a quartz BAW resonator coupled to a SQUID can be described by the following equation \cite{Goryachev2021},
\begin{equation}\label{eqn:Sh}
S_h^+(\omega) = \frac{\sqrt{S_u(\omega)}}{\left|H(i\omega)\right|}=\frac{\sqrt{S_v(\omega)}}{\kappa_\lambda \omega_\lambda Z_\mathrm{SQUID}(\omega)}\left|\frac{\frac{-\omega^2}{2}h_0\xi}{(i\omega)^2+\tau_\lambda^{-1}i\omega+\omega_\lambda^2}\right|^{-1},
\end{equation}
where $S_u(\omega)$ is the displacement spectral density of the resonator and has units m$^2\mathrm{Hz}^{-1}$, $S_v(\omega)$ is the spectral density of voltage fluctuations at the amplifier output with units V$^2\mathrm{Hz}^{-1}$, $\kappa_\lambda$ is an electromechanical coupling constant that relates charge on the BAW electrodes to displacement of the crystal \cite{Goryachev2014b}, $\omega_\lambda$ is the resonant frequency of mode $\lambda$, it's time constant is given by $\tau_\lambda = Q_\lambda/\omega_\lambda$ where $Q_\lambda$ is the acoustic quality factor, $h_0$ is the crystal longitudinal thickness and $\xi$ is a weighting term which parametrises the coupling of the BAW to gravitational waves. $H(i\omega)$ relates the displacement of the crystal to the characteristic strain of an impinging gravitational wave and thus has units of m$\left[\mathrm{strain}\right]^{-1}$.\\
It has previously been demonstrated \cite{Goryachev2014} that the effective mode temperature due to Nyquist noise fluctuations in a quartz resonator, while coupled to a SQUID, closely matches the physical temperature of the device. Here it was shown that based off the noise model of equation \ref{eqn:ZSQUID}, the energy stored in an acoustic mode of the resonator can be given by integrating the Lorentzian narrow-band noise:
\begin{equation}\label{eqn:Nyquist}
E = k_b T_\lambda = R_\lambda \frac{V_p^2}{4Z_\mathrm{SQUID}(\omega)^2} = \frac{\omega_\lambda M_\lambda}{\kappa_\lambda^2 Q_\lambda} \frac{V_p^2}{4Z_\mathrm{SQUID}(\omega)^2}.
\end{equation}
Where $k_b$ is Boltzmann's constant, $T_\lambda$ is the acoustic mode temperature, $M_\lambda$ is the effective mass, $V_p$ is the peak value of the Lorentzian seen in the voltage spectral density at the amplifiers output, and we have used the definition $\kappa_\lambda = \sqrt{\omega_\lambda M_\lambda/R_\lambda Q_\lambda}$. Thus, by setting $S_v(\omega)|_{\omega=\omega_\lambda} = V_p^2$, and substituting equation (\ref{eqn:Nyquist}) into (\ref{eqn:Sh}), the Nyquist noise produced by the quartz resonator in the vicinity of an acoustic mode can be related to the device's strain sensitivity, removing any dependence on the SQUID amplifier, 
\begin{equation}
S_h^+(\omega)\bigg\rvert_{\omega=\omega_\lambda} = ~\sqrt{\frac{4k_b T_\lambda \omega_\lambda}{Q_\lambda M_\lambda}}\left(\frac{1}{\omega_\lambda^2 h_0\xi}\right).
\end{equation}
The effective mass of an acoustic mode $M_\lambda$ can be determined analytically from the equations of linear piezoelectricity \cite{Tiersten1976}, $\xi$ is given by equation (12) of Goryachev \textit{et al}\cite{Goryachev2014b} where we assume well trapped phonons, and the longitudinal length $h_0$ for the quartz resonators utilised here is 0.5 mm. The intrinsic quality factor of the acoustic modes $Q_\lambda$ must be determined from experiment. 
Network analysis was thus employed in order to make accurate measurements of $Q_\lambda$, as this scheme allows for compensation of the effects of loading of the resonator by its coupling to an external environment. To this end, the quartz resonator in its copper enclosure was held at 4 K in a dilution refrigerator, as $Q_\lambda$ is highly temperature dependent, with each of its electrodes connected to a separate port of a vector network analyser. More details on such a measurement scheme can be found elsewhere \cite{Galliou2015, galliou:091911}. The quality factor was extracted by fitting to the measured transmission coefficient $S_{21}$, while using reflection coefficient measurements $S_{11}$ to determined any loading effects. An example of such a measurement of $S_{21}$ for a single mode is given in figure \ref{fig:Q}.\\
Three acoustic mode polarisations can be observed in these resonators; longitudinal (A) fast shear (B) and slow shear (C), we denote the mode $\lambda = X_{n,m,p}$ as the $n^\mathrm{th}$ overtone of the polarization family $X$, with higher order azimuthal variations dictated by the integers $n$ and $p$. The results of such measurements as well as the determined strain sensitivity are given in table \ref{tab:results} and plotted in figure \ref{fig:strain} .
We see from these results that the acoustic modes observed in this quartz resonator are capable of providing sensitivity to strain signals that is competitive with other HFGW experiments. While the sensitivity seemingly improves at higher frequencies due to improvements in resonator quality factor, this situation will reach a critical point when the coupling parameter $\xi$, as well as $M_\lambda$ starts to rapidly decrease for increasing $n$. In this work only acoustic modes with $m,p = 0$ where measured, where the bulk of the phonon distribution is centred in the crystal. There exists a large density of higher order $m,p \neq 0$ modes that could be monitored in order to sample more of the parameter space, however $\xi$ further decreases with increasing $m,p$. In order to further improve sensitivity to external strains, the device could be subject to colder dilution temperatures of $T~\approx$ 20 mK, reducing thermal vibration noise as well as increasing $Q_\lambda$ of some modes\cite{Screp}. However, the acquisition of new milli Kelvin SQUID sensors as well as investigations into the thermal state of a milli Kelvin quartz-SQUID coupled system, will need to be achieved before MAGE can be successfully operated at these lower temperatures. This is a potential upgrade for future iterations of MAGE.\\
\begin{figure}[ht]
\centering
\includegraphics[width = 0.48\textwidth]{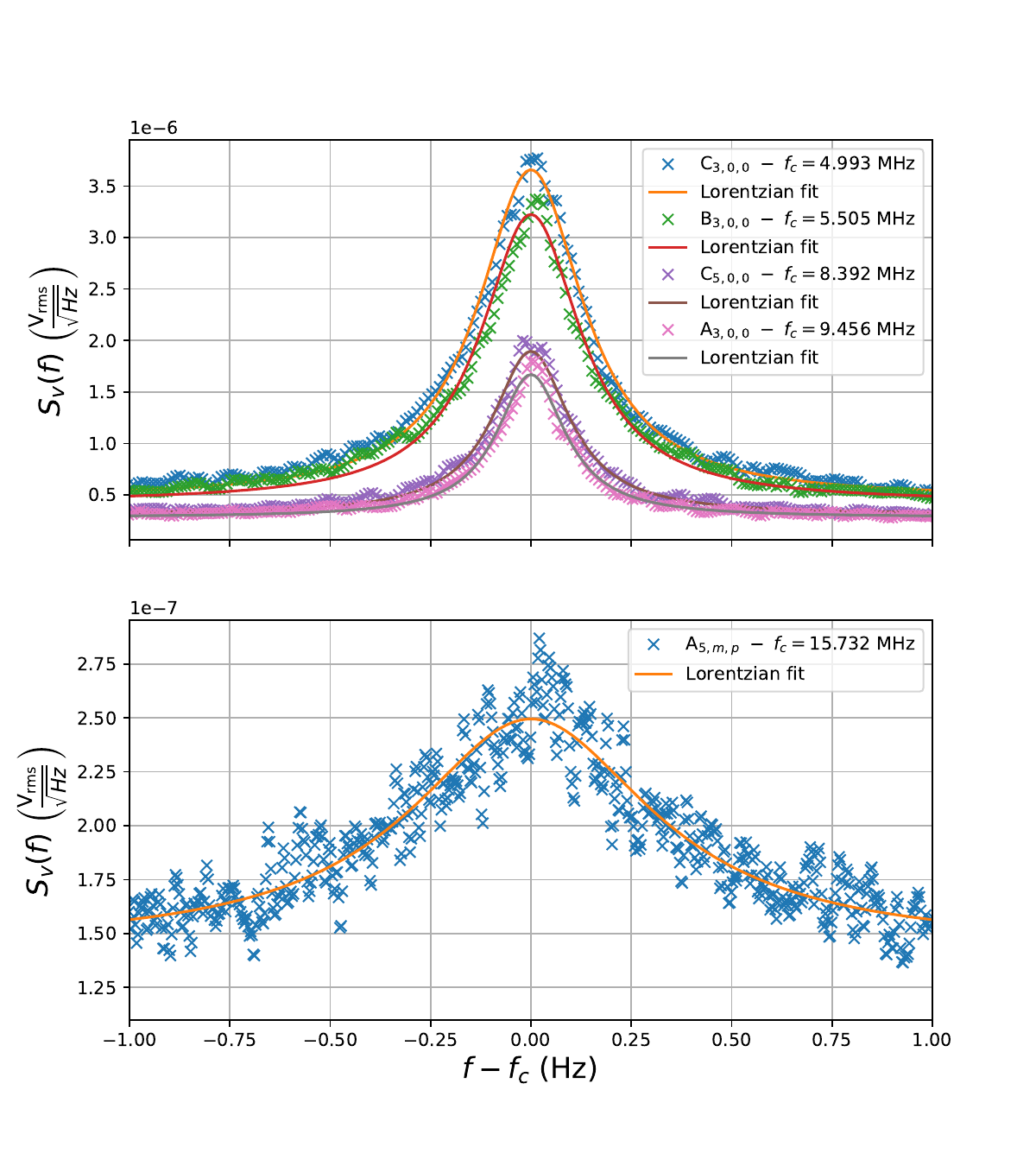}
\caption{\label{fig:MAGE0}Spectral density of voltage fluctuations of the amplifier output with the quartz resonator coupled to its input. Data of modes C$_{3,0,0}$, B$_{3,0,0}$, C$_{5,0,0}$, A$_{3,0,0}$ and A$_{5,0,0}$ is shown as well as the corresponding Lorentzian fits}
\end{figure}
MAGE will monitor a selection of acoustic modes including the ones presented here, simultaneously across both detectors at 4 K. The separation in frequency between the acoustic modes of the resonator measured here, and the same modes inherent to the resonator used in the previous GEN 1 and GEN 2 experiments, are listed in table \ref{tab:results} as $\Delta\omega$. This shows that a HFGW event signal with a frequency distribution of a few hundred Hz would be successfully observed in both detectors, however the modes remain too far separated to perform any cross-correlation analysis.\\
At these frequencies and for the strain sensitivity estimated, primordial black hole merger events would be the most motivated target HFGW source for MAGE. Based on the discussions presented in Aggarwal \textit{et al}\cite{Aggarwal2021}, and using the C$_{3,0,0}$ mode as an example, MAGE would be sensitive to a 0.0004 $M_\cdot$ black hole merger, where $M_\cdot$ denotes the solar mass. Such an event would produce a short burst of peak characteristic strain $h_c = 2.9\times10^{-17}$ incident upon the detector from a distance of 0.1 pc. The second detectors spectral strain sensitivity can be converted to dimensionless noise characteristic as $h_n = S^+_h(f)\sqrt{f} = 2.9\times10^{-17}$, putting the short burst of incident HFGW strain from such a merger event within the detectable capabilities of MAGE.\\
\begin{table*}[ht]
\centering
\begin{tabular}{lllllll}
\hline
\hline
$X_{n,m,p}$ & $\frac{\omega_\lambda}{2\pi}$(MHz) & $\frac{\Delta \omega}{2\pi}$(Hz) & $Q_\lambda$($10^7$) & $T_\lambda$ (K) & $S_h^+$ pred. ($10^{-20}$)& $S_h^+$ meas. ($10^{-20}$) \\
\hline
\hline
C$_{3,0,0}$ & 4.993 & 26.26 & 4.83 & 4.70 & 1.87 & 1.35\\
B$_{3,0,0}$ & 5.505 & 230.4 & 5.04 & 3.39 & 1.79 & 1.10\\
C$_{5,0,0}$ & 8.392 & 162.5 & 7.40 & 2.61 & 2.34 & -\\
B$_{5,0,0}$ & 9.152 & 106.2 & 9.66 & 2.69 & 1.98 & -\\
A$_{3,0,0}$ & 9.452 & 367.5 & 13.3 & 1.92 & 0.67 & -\\
\hline
\hline
\end{tabular}
\caption{\label{tab:results}This table summarises the main results of this work such as the measured quality factor $Q_\lambda$ and mode temperature $T_\lambda$, predicted strain sensitivity $S_h^+$, and measured strain sensitivity where applicable, for multiple modes of the resonator of interest. Also included is the separation in frequency $\Delta\omega$ between the modes of this resonator, and those of the one previously used for the GEN 1 and GEN 2 experiment\cite{Goryachev2021, Goryachev2014}.}
\end{table*}
\section*{Defector Thermal State}
With the SQUID amplifier calibrated, and the quartz resonator characterised, the vibrational mode temperature of the  assembled HFGW detector can be directly determined in order to reaffirm that the readout is indeed dominated by thermal Nyquist fluctuations, and that the calibration procedure and noise model presented in the preceding sections are accurate. Additionally, the ultimate usable bandwidth of the amplifier can be investigated by measuring the observed signal to noise ratio (SNR) of high frequency modes.\\
The output voltage spectral density of the amplifier was again monitored by a spectrum analyser. This set-up is identical to that of figure \ref{fig:Vnoise} b), with the exception of the load resistor being replaced by the quartz resonator. Multiple modes were observed in this output spectrum with well defined Lorentzian line shape, as seen in figure \ref{fig:MAGE0}. Their temperatures $T_\lambda$ can be determined by applying equation \ref{eqn:Nyquist}, the results of these calculations are given in table \ref{tab:results}. These results show that the system is well calibrated around the lower order modes where we see $T_\lambda$ close to the environment temperature, however this relationship begins to diverge for modes of higher order. This is due to the necessary re-tuning of the SQUID amplifier between experiments. It is necessary to re-tune the SQUID biasing and feedback parameters in between experiments as the parameters for optimal gain and bandwidth change upon thermal cycling of the operating environment. Variation of the SQUID gain due to such processes is the dominant source of error in calibrating total strain sensitivity. This uncertainty is much larger than those introduced by Lorentzian fitting or measurement, which provide less than $2\%$ error in the peak amplitude of the measured modes. The total uncertainty in the calibration procedure must thus be quantified as the difference between the measured and predicted strain sensitivities. In this particular case the device was not optimally tuned and suffered a reduced bandwidth, resulting in a deviation of the device's transimpedance from the calibrated value at high frequencies.\\
Future experiments will address this discrepancy by further fine tuning of the SQUID parameters, and re-performing the calibration procedure. However, for modes around the frequencies in which strong candidate events where seen in the GEN 1 and GEN 2 experiments, the current calibration is satisfactory. Thus the spectral strain sensitivity could be determined from the voltage spectra of modes C$_{3,0,0}$ and B$_{3,0,0}$ and compared to the predicted values. The results of these calculations are given in table \ref{tab:results}. Figure \ref{fig:MAGE0} further shows that although the amplifier bandwidth $f_\mathrm{3dB}$ was measured to be as low as 2.4 MHz, modes at frequencies as high as 15.732 MHz are still observable with 6 dB of SNR.
\section*{Conclusions}
In summary, a new gravitational wave detection experiment targeting the MHz frequency band has been introduced. MAGE will utilise two quartz BAW resonators coupled to SQUID amplifiers in order to transduce impinging strain fields into measurable voltage fluctuations. With two detectors in simultaneous operation, MAGE can exclude background events incident in just one detector from possible HFGW sources. The characteristics of the newly installed SQUID amplifier and quartz crystal resonator have been presented, such that the spectral strain sensitivity and vibrational mode temperature of this new detector could be determined.\\  
The current iteration of MAGE is actively taking data, monitoring the five modes presented in table \ref{tab:results} as well as the A$_{5,0,0}$ mode at 15.7 MHz. Although, in its current state, an additional 10 modes in each crystal could be observed by repeating the process presented in the sensitivity to HFGWs section. Immediate future work will see additional modes with $m,p \neq 0$ added to MAGEs current selection. The frequencies of these modes can be analytically determined and are easily found, although with the current hardware limited bandwidth, MAGE will be restricted to modes below $\approx 15.7$ MHz.\\
Further future improvements to MAGE will see the installation of a cosmic particle veto detector, an extension of MAGE's sensitivity to higher frequency quartz modes via hardware upgrade to the SQUID control electronics, and a possible push towards operation at milli-Kelvin temperatures. These current and planned developments provide the natural next step from the previous GEN 1 and GEN 2 experiment, which only featured a single quartz BAW detector. The immediate experimental program for MAGE will see it primarily target frequencies around ~ 5.5 MHz, where the GEN 1 and GEN 2 experiments detected strong signals. In order to further develop the quartz bulk acoustic wave technology for HFGW search experiments, it is critical the strong features seen in the GEN 1 and GEN 2 experiments are better understood. MAGE will hopefully be able to provide further insight into the origin of these transient features, maturing the quartz BAW architecture for future HFGW searches. MAGE is currently taking data intermittently as of early 2023. 
\section*{Data Availability}
The datasets generated during and/or analysed during the current study are available from the corresponding author on reasonable request
\bibliography{biblioMAGE}
\section*{Acknowledgements}
This research was supported by the Australian Research Council (ARC) Grant No. DP190100071, along with support from the ARC Centre of Excellence for Engineered Quantum Systems (EQUS, CE170100009) and the ARC Centre of Excellence for Dark Matter Particle Physics (CDM, CE200100008).\\
\section*{Author Contribution}
M.G. and M.T. conceived the idea presented. W.C. wrote the main manuscript text and performed the experimental work. All authors reviewed the manuscript.
\section*{Competing Interests}
The authors declare no competing interests

\end{document}